\PassOptionsToPackage{unicode}{hyperref}
\PassOptionsToPackage{hyphens}{url}
\PassOptionsToPackage{dvipsnames,svgnames,x11names}{xcolor}
\documentclass[
  english,
  runningheads]{llncs}
\usepackage{xcolor}
\usepackage{amsmath,amssymb}
\usepackage{iftex}
\ifPDFTeX
  \usepackage[T1]{fontenc}
  \usepackage[utf8]{inputenc}
  \usepackage{textcomp} 
\else 
  \usepackage{unicode-math} 
  \defaultfontfeatures{Scale=MatchLowercase}
  \defaultfontfeatures[\rmfamily]{Ligatures=TeX,Scale=1}
\fi
\usepackage[]{newtxtext}
\ifPDFTeX\else
\fi
\IfFileExists{upquote.sty}{\usepackage{upquote}}{}
\IfFileExists{microtype.sty}{
  \usepackage[]{microtype}
  \UseMicrotypeSet[protrusion]{basicmath} 
}{}
\makeatletter
\@ifundefined{KOMAClassName}{
  \IfFileExists{parskip.sty}{%
    \usepackage{parskip}
  }{
    \setlength{\parindent}{0pt}
    \setlength{\parskip}{6pt plus 2pt minus 1pt}}
}{
  \KOMAoptions{parskip=half}}
\makeatother
\makeatletter
\ifx\paragraph\undefined\else
  \let\oldparagraph\paragraph
  \renewcommand{\paragraph}{
    \@ifstar
      \xxxParagraphStar
      \xxxParagraphNoStar
  }
  \newcommand{\xxxParagraphStar}[1]{\oldparagraph*{#1}\mbox{}}
  \newcommand{\xxxParagraphNoStar}[1]{\oldparagraph{#1}\mbox{}}
\fi
\ifx\subparagraph\undefined\else
  \let\oldsubparagraph\subparagraph
  \renewcommand{\subparagraph}{
    \@ifstar
      \xxxSubParagraphStar
      \xxxSubParagraphNoStar
  }
  \newcommand{\xxxSubParagraphStar}[1]{\oldsubparagraph*{#1}\mbox{}}
  \newcommand{\xxxSubParagraphNoStar}[1]{\oldsubparagraph{#1}\mbox{}}
\fi
\makeatother

\usepackage{longtable,booktabs,array}
\usepackage{calc} 
\usepackage{etoolbox}
\makeatletter
\patchcmd\longtable{\par}{\if@noskipsec\mbox{}\fi\par}{}{}
\makeatother
\IfFileExists{footnotehyper.sty}{\usepackage{footnotehyper}}{\usepackage{footnote}}
\makesavenoteenv{longtable}
\usepackage{graphicx}
\makeatletter
\newsavebox\pandoc@box
\newcommand*\pandocbounded[1]{
  \sbox\pandoc@box{#1}%
  \Gscale@div\@tempa{\textheight}{\dimexpr\ht\pandoc@box+\dp\pandoc@box\relax}%
  \Gscale@div\@tempb{\linewidth}{\wd\pandoc@box}%
  \ifdim\@tempb\p@<\@tempa\p@\let\@tempa\@tempb\fi
  \ifdim\@tempa\p@<\p@\scalebox{\@tempa}{\usebox\pandoc@box}%
  \else\usebox{\pandoc@box}%
  \fi%
}
\def\fps@figure{htbp}
\makeatother

\ifLuaTeX
\usepackage[bidi=basic,shorthands=off]{babel}
\else
\usepackage[bidi=default,shorthands=off]{babel}
\fi
\ifLuaTeX
  \usepackage{selnolig} 
\fi

\usepackage[numbers,sectionbib]{natbib}
\usepackage{booktabs}
\usepackage{longtable}
\usepackage{array}
\usepackage{multirow}
\usepackage{wrapfig}
\usepackage{float}
\usepackage{colortbl}
\usepackage{pdflscape}
\usepackage{tabu}
\usepackage{threeparttable}
\usepackage{threeparttablex}
\usepackage[normalem]{ulem}
\usepackage{makecell}
\usepackage{xcolor}
\usepackage[varvw]{newtxmath}
\usepackage{graphicx}
\usepackage{booktabs}
\usepackage{longtable}
\usepackage{array}
\usepackage{multirow}
\usepackage{makecell}
\usepackage{float}
\usepackage{adjustbox}
\usepackage{url}

\makeatletter
\@ifpackageloaded{caption}{}{\usepackage{caption}}
\AtBeginDocument{%
\ifdefined\contentsname
  \renewcommand*\contentsname{Table of contents}
\else
  \newcommand\contentsname{Table of contents}
\fi
\ifdefined\listfigurename
  \renewcommand*\listfigurename{List of Figures}
\else
  \newcommand\listfigurename{List of Figures}
\fi
\ifdefined\listtablename
  \renewcommand*\listtablename{List of Tables}
\else
  \newcommand\listtablename{List of Tables}
\fi
\ifdefined\figurename
  \renewcommand*\figurename{Figure}
\else
  \newcommand\figurename{Figure}
\fi
\ifdefined\tablename
  \renewcommand*\tablename{Table}
\else
  \newcommand\tablename{Table}
\fi
}
\@ifpackageloaded{float}{}{\usepackage{float}}
\floatstyle{ruled}
\@ifundefined{c@chapter}{\newfloat{codelisting}{h}{lop}}{\newfloat{codelisting}{h}{lop}[chapter]}
\floatname{codelisting}{Listing}

\makeatother
\makeatletter
\@ifpackageloaded{caption}{}{\usepackage{caption}}
\@ifpackageloaded{subcaption}{}{\usepackage{subcaption}}
\makeatother
\usepackage{bookmark}
\IfFileExists{xurl.sty}{\usepackage{xurl}}{} 
\makeatletter
\@ifundefined{xmpquote}{}{}
\makeatother
\hypersetup{
  pdflang={en},
  colorlinks=true,
  linkcolor={blue},
  filecolor={Maroon},
  citecolor={Blue},
  urlcolor={Blue},
  pdfcreator={LaTeX via pandoc}}

\author{}
\date{}
\begin{document}

\title{Same Performance, Different Process: Epistemic Ownership in AI-Mediated Education}
\subtitle{PRE-PRINT, v.2026\_30\_09}
\titlerunning{Same Performance, Different Process}
\author{Jorge Fábrega}
\authorrunning{J. Fábrega}
\institute{CICS-UDD, Chile}
\maketitle

\begin{abstract}
Generative AI weakens the link between assessed performance and the cognitive processes that educational work is expected to reflect. This paper examines observable traces of those processes in a sample of 150 student–AI conversations. The analysis uses the concept of epistemic ownership as an interpretive framework, focusing on three dimensions that can become visible in interaction: Direction, Integration, and Evaluation. The results show that similar assessed performance can coexist with substantially different observable patterns of cognitive participation. Assignment scores therefore provide information about assessed outcomes but do not recover how cognitive work was distributed between student and AI during the interaction. The study provides an empirical basis for asking how assessment should incorporate evidence about cognitive process, rather than relying exclusively on the quality of the resulting product.
\end{abstract}

\section{Introduction}\label{introduction}

Generative artificial intelligence creates a distinctive problem for
education because it weakens a relationship between performance and
learning on which educational practice has long relied. Teachers
routinely have inferred something about learning from the work students
produce. That inference has always been imperfect, but an essay or a
solved problem was considered good enough evidence that the student had
engaged with the material and organized it into an intelligible result.
The output therefore provided a sufficiently informative signal of the
learning process. That can no longer be taken for granted.

Generative AI changes that evidentiary relation by entering activities
that education often treats as formative rather than merely
instrumental. Large language models can formulate research questions,
propose interpretations, and produce lines of reasoning that the user
had not previously articulated or only superficially understood. This
capacity makes the educational effects of generative AI increasingly
contingent on how students use the system and on the learning activity
in which that use is embedded
\citep{kasneci2023chatgpt, deng2025chatgpt, ma2025meta}. Now, a student
can produce a fluent explanation or polished technical artifact while
externalizing substantial portions of the work that such outputs
previously signaled without having demonstrated the cognitive effort
that was expected from the assignment given to him.

Therefore, there is a potential new kind of gap between performance and
learning when the learner delegates to a generative AI system cognitive
processes that contribute directly to the competence being formed.
Searching for a typo or reorganizing references usually occupies a
limited role in disciplinary judgment, but formulating a problem or
deciding why one explanation is preferable to another occupies a
different position. Learners develop these capacities through repeated
exercise, often under conditions of uncertainty, and instructional
designs that preserve cognitively productive difficulty can improve
subsequent learning \citep{kapur2008productive}.

Education therefore faces a problem of cognitive distribution.
Generative AI can remove low-value effort and increase access to
explanation, but it can also perform operations that learners would
otherwise have to formulate, inspect, revise, or connect themselves.
These operations matter because learning depends partly on what learners
are cognitively required to do with information rather than on exposure
to a finished result alone. Recent experimental work similarly shows
that stronger task performance with generative AI can coexist with
weaker gains in knowledge or transfer when AI use reduces metacognitive
engagement \citep{fan2025metacognitive}.

The learning process itself is only partially observable. This paper
focuses on one window into that process: the recorded interaction
through which a learner delegates cognitive work to a generative AI
system and subsequently responds to what the system produces. Such
interactions do not reveal everything the learner thinks, understands,
or does outside the conversation. They can, however, provide evidence
about how cognitive work is distributed within the AI-mediated activity:
whether the learner governs its direction, incorporates contributions
produced by the system, and participates in judging the resulting work.

The paper uses epistemic ownership to analyze this relation
\citep{fabrega2026epistemic}. Epistemic ownership concerns the learner's
justificatory command over work produced through distributed cognition
and comprises four capacities: direction, integration, evaluation and
answerability. The empirical analysis concentrates on observable
interactional manifestations of the first three. Answerability requires
an additional opportunity for the learner to reconstruct and defend the
reasoning embodied in the resulting work. The central empirical question
is therefore whether assessed performance alone distinguishes
interactions in which these forms of learner participation become
differently observable.

\section{Theoretical framework: epistemic ownership under cognitive
delegation}\label{theoretical-framework-epistemic-ownership-under-cognitive-delegation}

Learning has always relied on external cognitive resources. Diagrams,
calculators, databases, and other artifacts redistribute cognitive work
across learners and their environments
\citep{clark1998extended, risko2016cognitive}. Their educational
significance depends on how that redistribution relates to the capacity
an activity is intended to develop. Generative AI expands this problem
because it can contribute information and computation, as well as
explanations, interpretations, arguments, and lines of reasoning.

The relevant educational question is therefore the learner's command
over work produced through this distributed process. Epistemic ownership
captures that relation. A learner possesses stronger epistemic ownership
when she can reconstruct and defend the reasoning embodied in the
resulting work. Substantial AI assistance can coexist with such command
when the learner retains a clear purpose, incorporates contributions
into her own reasoning, and judges whether they deserve acceptance.
Strong performance may also coexist with weak command when an adequate
AI contribution is adopted and carried forward without the learner
having to reconstruct the reasoning it contains. This condition is
referred to as justificatory dispossession \citep{fabrega2026epistemic}.

Epistemic ownership comprises four interdependent functions: direction,
integration, evaluation, and answerability \citep{fabrega2026epistemic}.
The present analysis focuses on the first three because they can leave
observable traces during AI-mediated work.

Direction concerns the learner's control over the purpose and trajectory
of the activity. It becomes visible when the student specifies relevant
constraints, selects or changes a procedure, or redirects the
interaction in light of the task. Such control also has a metacognitive
component because decisions about when and how to use external
assistance depend partly on judgments about effort and expected
performance \citep{dunn2016metacognitive}.

Integration concerns how AI contributions enter the learner's subsequent
reasoning or action. It becomes observable when the student applies,
transforms, or otherwise carries forward a prior AI contribution. Mere
topical continuity or the appearance of AI-generated material in a final
artifact does not establish integration.

Evaluation concerns the learner's participation in judging whether an
external contribution should be accepted, revised, or rejected. It
becomes visible through moves such as testing a solution, detecting a
discrepancy, applying a criterion, or correcting a procedure. This
dimension connects epistemic ownership to research on epistemic
cognition, where standards for evidence and justification shape
judgments about knowledge claims \citep{chinn2011epistemic}. Asking the
AI to perform the judgment does not, by itself, constitute learner
evaluation.

The fourth one, answerability, concerns whether the learner can
reconstruct and defend the reasoning embodied in the final work. It
differs evidentially from the preceding dimensions because an
interaction record may never require the student to explain or defend
that reasoning. Establishing answerability therefore requires additional
forms of elicitation and lies beyond the empirical analysis developed
here.

These distinctions matter for assessment. Two students may submit
similarly successful work while having exercised different degrees and
forms of direction, integration, and evaluation. AI-mediated education
therefore creates an observability problem: educators need evidence
about the processes through which relevant capacities were exercised,
particularly when the system can perform operations that the assignment
is intended to develop \citep{ajjawi2024authenticity}. Interaction
histories provide one such source of evidence because they preserve
parts of the sequence through which work was delegated, resumed,
modified, tested, or redirected.

\subsection{From epistemic ownership to observable interactional
evidence}\label{from-epistemic-ownership-to-observable-interactional-evidence}

Epistemic ownership itself cannot be read directly from a chat
transcript. What the transcript can reveal are observable actions that
bear on its component functions. The analytical task is therefore to
identify evidence of direction, integration, and evaluation in the
interaction while remaining explicit about the limits of those
inferences.

Human-AI interaction is useful for this purpose because it unfolds
sequentially. A student contributes, the system responds, and the
student may then continue, redirect, apply, modify, question, test,
reject, or abandon what the system has produced. The epistemic
significance of a learner move therefore depends partly on what preceded
it and what follows. This principle draws selectively on conversation
analysis and related work on grounding and repair, which treat
subsequent action as evidence of how a prior contribution has been
understood and taken up
\citep{schegloff1968, schegloff2007sequence, clark1989, schegloff1977}.
Similar relational patterns can be examined in interactions with
conversational systems and large language models, although human-AI chat
does not reproduce the social conditions of ordinary conversation
\citep{porcheron2018, balaraman2023thirdpositionrepair, shaikh2025, lachenmaier2026}.

For the present study, human re-entry after an AI contribution is
therefore a particularly informative locus. Direction is coded when the
learner exercises task-relevant control over the purpose, constraints,
procedure, artifact, or trajectory of the activity. Integration is coded
when a subsequent learner action can be specifically connected to the
uptake, application, transformation, or reuse of a prior AI
contribution. Evaluation is coded when the learner participates in
judging a contribution, procedure, artifact, or result through
inspection, comparison, testing, discrepancy detection, diagnosis,
correction, or criterion-based judgment. These dimensions are
analytically independent and may overlap within the same learner move.

Interpretation remains sequence-dependent. A constraint may originate in
the assignment, the learner, a prior AI suggestion, or the working
environment, and that provenance affects what can be inferred from its
use. Likewise, returning with the output of AI-generated code can
provide evidence of integration, while identifying why that output is
incorrect can also provide evidence of evaluation. The student turn
supplies the immediate evidence, the surrounding sequence establishes
what it responds to, and the conversation trajectory shows how these
forms of participation are organized over time.

The resulting codes describe observable participation in the AI-mediated
process. They do not constitute an additive measure of epistemic
ownership, and absence of a coded behavior does not establish absence of
the underlying capacity. Evaluation, direction, or integration may occur
outside the chat and leave no trace in the interaction record. The
empirical analysis therefore uses these dimensions to characterize
observable architectures of cognitive participation and to examine how
they relate to performance, while reserving broader claims about
epistemic ownership for evidence that the available data can support.

\section{Data and Method}\label{data-and-method}

\subsection{StudyChat data}\label{studychat-data}

The empirical analysis uses the StudyChat dataset
\citep{mcnichols2026studychat}, which records student interactions with
an LLM-based chatbot during Fall 2024 and Spring 2025. Each record
contains a student prompt and the corresponding LLM response, together
with conversation (\texttt{chatId}), student (\texttt{userId}),
assignment (\texttt{topic}), and interaction-order
(\texttt{interactionCount}) identifiers.

The corpus contains 2,214 conversations from 203 students completing 7
course assignments in artificial intelligence and computer science,
including programming, data preprocessing, regression and
classification, and language-modeling tasks. In total, the conversations
comprise 16,851 student utterances and corresponding LLM responses, with
a mean of 7.6 student turns per conversation (median = 4). The
conversation is the sampling unit, while individual student turns
provide the within-conversation evidence used to identify Direction,
Integration, and Evaluation. The reconstructed corpus retains semester,
student, assignment, and interaction-order identifiers, allowing
conversations to be sampled across these dimensions and checked for
ordering and identifier consistency.

\subsection{Coding observable cognitive
participation}\label{coding-observable-cognitive-participation}

The coding scheme provides an exploratory operationalization of
observable manifestations of Direction, Integration, and Evaluation. The
student turn is the coding unit, interpreted in relation to its local
sequence and the conversation as a whole. The dimensions are multilabel
and analytically independent and are therefore not combined into an
overall epistemic-ownership score. Additional fields record how a turn
relates to prior AI output, the provenance of relevant content or
criteria, validation and diagnostic moves, and the evidence supporting
each code.

The coding protocol was developed through iterative human-in-the-loop
analysis of 40 heterogeneous StudyChat conversations. The process
focused on recurring boundary cases among the three dimensions and
reached a practical saturation point after 23 cases, when additional
conversations no longer required new recurring distinctions or boundary
rules. The resulting protocol was then tested on an independent
14-conversation pilot: six cases provided a technical test of the
prompt, JSON schema, API pipeline, and semantic validation, followed by
eight additional cases for methodological calibration. The codebook,
prompt, and output schema were subsequently frozen before application to
the main analytical sample of 150 conversations.

These methodological decisions were translated into operational
boundaries for identifying Direction, Integration, and Evaluation in
student--AI interactions. Table 1 summarizes the question guiding each
dimension, the forms of positive evidence considered, and the main
boundary used to distinguish genuine evidence from superficially similar
moves.

\begin{table}[H]
\centering
\caption{\label{tab:operational-rules}Operational boundaries for the three observable dimensions}
\centering
\fontsize{9}{11}\selectfont
\begin{tabular}[t]{>{\raggedright\arraybackslash}p{1.58cm}>{\raggedright\arraybackslash}p{3.05cm}>{\raggedright\arraybackslash}p{2.94cm}>{\raggedright\arraybackslash}p{2.94cm}}
\toprule
Dimension & Operational\_question & Positive\_evidence & Key\_boundary\\
\midrule
Direction & Does the student observably govern the task, assistance, procedure, artifact, constraint, decision, or trajectory? & Specification, constraint, procedural choice, redirection, correction of course, or control of workflow state & Speaking, asking any question, or initiating a conversation is insufficient by itself\\
Integration & Does the student observably incorporate, reuse, transform, apply, or transfer a prior AI contribution? & Returned use of AI-derived code or concepts, selective adoption, transformation, execution, reuse, or transfer & Sequential continuity or repeated requests to revise an AI artifact are insufficient without observable uptake\\
Evaluation & Does the student inspect, judge, compare, challenge, verify, or diagnose using an observable criterion? & Correction, discrepancy, comparison, hypothesis, test, diagnostic claim, or criterion-based judgment & Asking the AI 'is this correct?' is verification-seeking unless the student also contributes a criterion or judgment\\
\bottomrule
\end{tabular}
\end{table}

Two coding rules were central. Evidence was classified as explicit when
directly observable and as inferable only when the sequence supported a
specific inference over plausible alternatives. Provenance was recorded
separately for content, criteria, and judgments, allowing assignment
criteria, AI-derived material, and runtime feedback to contribute
differently to Direction, Integration, and Evaluation without being
treated as equivalent evidence.

The protocol was applied to 150 conversations sampled proportionally
across semester, assignment, and conversation-length strata, with short,
medium, and long bands defined by the 25th and 75th percentiles. Coding
used the OpenAI Responses API with GPT-5.6 Sol (reasoning\_effort =
``medium'') and Structured Outputs constrained by the schema. Each
conversation was coded from its full transcript, producing turn-level
codes, evidence records, episode segmentation, and conversation
trajectories, with semantic validation and manual supervision of a
random subsample.

Because the dimensions are analytically independent, a learner move may
provide evidence of more than one dimension. Conversely, absence of
Integration or Evaluation indicates only that these forms of
participation were not observed in the recorded interaction, especially
when short conversations provide few opportunities for learner re-entry.

\subsection{Observable process in a task environment with immediate
feedback}\label{observable-process-in-a-task-environment-with-immediate-feedback}

StudyChat offers an unusually informative setting for examining the
observable process of human-AI cognitive delegation. The course tasks
are centered on computer science and data-analysis problems. Students
frequently move between the conversation and an executable environment,
where code runs, fails, produces an unexpected output, or returns a
concrete result. These external consequences create repeated
opportunities for the student to specify what should happen next,
implement a prior AI contribution, inspect a result, diagnose a failure,
and return to the model with new evidence.

This task ecology matters for interpretation. Direction, Integration,
and Evaluation should be especially observable when the activity
produces rapid external feedback (as in this case). A programming error,
a traceback, a changed coefficient, or a successful execution can become
part of the next student turn. In activities whose consequences are less
immediate, cognitive work may leave fewer traces in the chat.

Applying the coding protocol to the 150-conversation sample produced the
distribution shown in the table below. Because Integration and
Evaluation require opportunities for student re-entry, the results are
also reported across alternative minimum-turn thresholds.

\begin{longtable}[t]{rrrrrr}
\caption{\label{tab:table-process-observability}Observable process dimensions by available conversation length}\\
\toprule
Minimum turns & Conversations & Direction & Integration & Evaluation & D + I + E\\
\midrule
All & 150 & 146 (97.3\%) & 56 (37.3\%) & 69 (46.0\%) & 45 (30.0\%)\\
5 & 65 & 65 (100.0\%) & 43 (66.2\%) & 44 (67.7\%) & 35 (53.8\%)\\
8 & 48 & 48 (100.0\%) & 36 (75.0\%) & 37 (77.1\%) & 31 (64.6\%)\\
10 & 38 & 38 (100.0\%) & 31 (81.6\%) & 31 (81.6\%) & 27 (71.1\%)\\
12 & 29 & 29 (100.0\%) & 24 (82.8\%) & 25 (86.2\%) & 21 (72.4\%)\\
\addlinespace
15 & 21 & 21 (100.0\%) & 19 (90.5\%) & 19 (90.5\%) & 17 (81.0\%)\\
\bottomrule
\end{longtable}

Direction is visible in 146 of the 150 sampled conversations (97.3\%).
Once the trace contains at least five student turns, Direction is
present in every conversation in the current sample. Integration and
Evaluation show a different pattern. Both become progressively more
observable as conversation length increases, rising from 37.3\% and
46.0\%, respectively, in the full sample to 90.5\% and 90.5\% among
conversations with at least 15 student turns. This pattern is consistent
with the sequential nature of these dimensions: while Direction can be
expressed in an initial request or constraint, Integration and
Evaluation generally require subsequent opportunities for the learner to
respond to, apply, inspect, or judge an AI contribution.

\subsection{Linking observable process to assessed
performance}\label{linking-observable-process-to-assessed-performance}

The score data released with StudyChat report students' normalized
course-assignment grades. Each score corresponds to one of the seven
graded assignments and is expressed on a normalized scale ranging
approximately from 0 to 1 \citep{mcnichols2026studychat}. We use these
scores as measures of assessed performance under the grading criteria of
the corresponding assignment. They are not treated as direct measures of
learning, cognitive effort, or epistemic ownership. The empirical
question is whether similar performance signals coexist with different
observable configurations of Direction, Integration, and Evaluation.

The score linkage uses exact semester, student, and assignment
identifiers. Scores are standardized within each semester-by-assignment
cell using the complete released grade files before restricting the data
to the analytical sample. Zeros are retained as observed scores and no
value is clipped. At lower conversation-length thresholds, where more
than one sampled conversation can correspond to the same
student-assignment pair, the analysis retains the longest eligible
conversation and breaks ties by \texttt{case\_id}. The selection rule
uses neither the score nor the D/I/E coding.

Exact score linkage was available for 134 of the 150 conversations.
Unmatched cases were excluded from the performance comparisons; no
approximate linkage or imputation was used.

\begin{table}[H]
\centering
\caption{\label{tab:table-score-sample-flow}Analytical sample across alternative opportunity thresholds}
\centering
\fontsize{9}{11}\selectfont
\begin{tabular}[t]{>{\raggedleft\arraybackslash}p{1.11cm}>{\raggedleft\arraybackslash}p{1.11cm}>{\raggedleft\arraybackslash}p{1.11cm}>{\raggedleft\arraybackslash}p{1.39cm}>{\raggedleft\arraybackslash}p{1.66cm}>{\raggedleft\arraybackslash}p{1.29cm}>{\raggedleft\arraybackslash}p{1.57cm}}
\toprule
Minimum turns & Selected pairs & Scored pairs & All 3 observed & >=1 not observed & All 3 explicit & >=1 not explicit\\
\midrule
5 & 62 & 59 & 32 & 27 & 19 & 40\\
8 & 47 & 45 & 29 & 16 & 19 & 26\\
10 & 38 & 36 & 25 & 11 & 17 & 19\\
12 & 29 & 28 & 20 & 8 & 15 & 13\\
15 & 21 & 20 & 16 & 4 & 14 & 6\\
\bottomrule
\end{tabular}
\end{table}

To examine how observability changes as interaction length increases, we
use several minimum-turn thresholds. These thresholds provide a simple
way to compare conversations with different amounts of available
interaction and to observe whether the patterns in Integration and
Evaluation change as more opportunities for learner re-entry become
available. They are used descriptively rather than as substantively
meaningful cutoffs.

\section{Results}\label{results}

The first comparison distinguishes conversations in which all three
observable dimensions---Direction, Integration, and Evaluation---are
observed at least once from those in which at least one dimension is not
observed, allowing both explicit and conservatively inferable evidence.
The second uses a stricter evidentiary criterion, requiring at least one
explicit evidence record for each dimension. These groupings summarize
patterns of observable D/I/E coverage within the interaction record.
They are not intended as a scale, an overall measure of epistemic
ownership, or a classification of the underlying cognitive process.

\begin{table}[H]
\centering
\caption{\label{tab:table-threshold-contrasts}Descriptive performance contrasts across opportunity thresholds. Differences are all three D/I/E dimensions observed minus at least one not observed.}
\centering
\fontsize{9}{11}\selectfont
\begin{tabular}[t]{>{\raggedleft\arraybackslash}p{0.83cm}>{\raggedright\arraybackslash}p{2.13cm}>{\raggedleft\arraybackslash}p{1.29cm}>{\raggedleft\arraybackslash}p{1.25cm}>{\raggedleft\arraybackslash}p{1.25cm}>{\raggedleft\arraybackslash}p{1.25cm}>{\raggedleft\arraybackslash}p{1.25cm}}
\toprule
Min. turns & Definition & N (all 3 / >=1 absent) & Mean score diff. & Median score diff. & Mean z diff. & Median z diff.\\
\midrule
5 & Explicit evidence only & 19 / 40 & 0.0144 & 0.0107 & -0.0454 & -0.0694\\
5 & Observable evidence & 32 / 27 & 0.0040 & -0.0224 & -0.0927 & -0.1222\\
8 & Explicit evidence only & 19 / 26 & 0.0181 & 0.0141 & 0.0111 & -0.0694\\
8 & Observable evidence & 29 / 16 & 0.0094 & -0.0308 & -0.0650 & -0.1138\\
10 & Explicit evidence only & 17 / 19 & 0.0308 & 0.0242 & 0.1403 & 0.3088\\
10 & Observable evidence & 25 / 11 & 0.0282 & -0.0036 & 0.0628 & -0.1133\\
12 & Explicit evidence only & 15 / 13 & 0.0224 & 0.0157 & 0.1204 & 0.3474\\
12 & Observable evidence & 20 / 8 & 0.0219 & -0.0304 & 0.0393 & -0.1627\\
15 & Explicit evidence only & 14 / 6 & 0.0412 & 0.0271 & 0.3209 & 0.1528\\
15 & Observable evidence & 16 / 4 & 0.0405 & -0.0161 & 0.1119 & -0.2079\\
\bottomrule
\end{tabular}
\end{table}

Among conversations with linked assignment scores, the median assignment
score is 0.980 (IQR = 0.073), and 41.0\% have a score of exactly 1.00.
Across the alternative opportunity thresholds, the score comparisons do
not reveal a stable performance difference between conversations in
which all three dimensions are observed and those in which at least one
is not. Mean and median contrasts remain small in the original score
scale and vary in magnitude and, in some cases, direction as the minimum
interaction threshold changes. There is therefore little in these
descriptive comparisons to suggest a systematic score difference that
would warrant a stronger performance-based interpretation.

This is the relevant result for the argument developed here. Similar
assessed performance can accompany different observable patterns of
Direction, Integration, and Evaluation. The assignment score therefore
does not allow the observable process that produced that performance to
be reconstructed, while the interaction record provides evidence of
meaningful variation in at least these three dimensions of epistemic
ownership. Notice though that the sample is not large enough to support
population-level claims.

\subsection{Same performance signal, different observable
process}\label{same-performance-signal-different-observable-process}

The most direct illustration comes from comparisons within the same
semester and assignment. We selected within-semester, within-assignment
pairs with identical or closely matched scores, contrasting observable
D/I/E configurations, and sufficient interaction to make those contrasts
interpretable. These comparisons do not establish causal equivalence.
They identify concrete cases in which comparable assessed performance
coexists with different observable patterns of participation in the
epistemic process.

\begin{table}[H]
\centering
\caption{\label{tab:selected-pair-table}Selected within-cell pairs for qualitative reconstruction}
\centering
\fontsize{9}{11}\selectfont
\begin{tabular}[t]{>{\raggedleft\arraybackslash}p{0.65cm}>{\raggedright\arraybackslash}p{1.11cm}>{\centering\arraybackslash}p{0.65cm}>{\raggedright\arraybackslash}p{4.44cm}>{\raggedleft\arraybackslash}p{0.74cm}>{\centering\arraybackslash}p{0.74cm}>{\raggedleft\arraybackslash}p{0.92cm}}
\toprule
Pair & Context & Case & Case ID & Turns & D/I/E & Score\\
\midrule
1 & S25-a4 & A & S25::74d95a7d-3535-4768-abab-884c6c027b76 & 27 & 1/1/1 & 1.0000\\
1 & S25-a4 & B & S25::ff387f68-9975-4a8c-a8d6-0cc746503425 & 17 & 1/1/0 & 1.0000\\
2 & F24-a3 & A & F24::f0509d86-59f3-4b08-a9a8-4342e3e703d2 & 6 & 1/1/1 & 0.8727\\
2 & F24-a3 & B & F24::14eee44b-f0e8-4665-b62f-787676d60c33 & 6 & 1/0/0 & 0.8727\\
\bottomrule
\end{tabular}
\end{table}

These comparisons make the central empirical point in its most concrete
form: the performance signal does not uniquely identify the observable
cognitive process associated with it. The dialogue traces add
information that the score alone does not contain.

\subsection{Qualitative reconstruction of two matched
pairs}\label{qualitative-reconstruction-of-two-matched-pairs}

The descriptive comparison can be made more concrete by examining
selected moments from two pairs of conversations with identical
assignment scores. The excerpts below do not reproduce the full
interactions. They select moments that make visible how students
directed the task, incorporated prior AI contributions, and participated
- or did not participate - in evaluating the resulting work.

\emph{Pair 1. Same score and observable Integration, different
Evaluation}

Both students received a score of 1.00 in S25 Assignment 4. Both
conversations contain sustained Direction and observable Integration.
Their trajectories differ in whether the student participates in
Evaluation.

\begin{table}[H]

\caption{\label{tbl-pair1-process}Selected process evidence for Pair 1}

\centering{

\centering\begingroup\fontsize{9}{11}\selectfont

\begin{tabular}{>{\raggedright\arraybackslash}p{2.19cm}>{\raggedright\arraybackslash}p{4.37cm}>{\raggedright\arraybackslash}p{4.37cm}}
\toprule
Process moment & Conversation A: D + I + E, 27 turns & Conversation B: D + I, no E, 17 turns\\
\midrule
Direction / criterion setting & “for 3 did you use these instructions: … Report the score … explain what the score means … Extract the coefficients and intercept…” The student does not merely request completion. The prior response is checked against specific assignment requirements. & “X = slime[‘Temperature °C’, ‘Mols KCL’] … gets keyerror, why?” The student identifies the task and failure but delegates explanation of the problem to the AI.\\
Integration / re-entry after using AI assistance & “why did I get this warning: … X does not have valid feature names, but LinearRegression was fitted with feature names.” The student has executed the preceding workflow and returns with its concrete consequence. & “make this a copyable table: [Logistic Regression 0.856559 … SVM 0.928172 …]” The returned model results are consistent with prior AI-supported procedures, providing evidence that those procedures entered the student's subsequent work.\\
Evaluation & After executing a revised neural-network workflow, the student returns to the generated artifact and instructs: “do not print every iteration.” Here the student supplies the judgment and the desired correction rather than asking the AI whether the result is acceptable. & No comparable move appears across the 17 student turns. Results are returned and further assistance is requested, but the student does not articulate a correctness criterion, comparative judgment, diagnostic hypothesis, or interpretation of model performance.\\
\bottomrule
\end{tabular}
\endgroup{}

}

\end{table}%

The relational basis of these codes is visible in a short sequence. The
AI proposed an \texttt{MLPClassifier} workflow using
\texttt{verbose=True} to monitor training progress. The student
subsequently returned the same workflow and added: ``do not print every
iteration.'' The re-entry shows both uptake of the AI-produced artifact
and an explicit judgment about its behavior: the prior contribution has
entered the student's work, but it is not simply accepted unchanged. In
Conversation B, re-entry is also present, including the return of model
results, but it remains oriented toward obtaining the next
transformation or explanation. The coding therefore identifies
Integration but no Evaluation across the interaction.

\emph{Pair 2. Same score and interaction length, different Integration
and Evaluation}

The second pair provides a broader contrast under closely matched
conditions. Both students worked on F24 Assignment 3, received the same
normalized assignment score of 0.8727, and interacted with the AI for
six student turns. Conversation A contains observable Direction,
Integration, and Evaluation. Conversation B contains Direction, but
neither Integration nor Evaluation is observed.

\begin{table}[H]

\caption{\label{tbl-pair2-process}Selected process evidence for Pair 2}

\centering{

\centering\begingroup\fontsize{9}{11}\selectfont

\begin{tabular}{>{\raggedright\arraybackslash}p{2.19cm}>{\raggedright\arraybackslash}p{4.37cm}>{\raggedright\arraybackslash}p{4.37cm}}
\toprule
Process moment & Conversation A: D + I + E, 6 turns & Conversation B: D only, 6 turns\\
\midrule
Direction & The student advances through concrete assignment tasks, including data export, preprocessing, and SQL operations, specifying what should be done at each stage. & The student also directs the interaction through successive assignment-related requests concerning unique\_ID, Markdown conversion, variable renaming, SQL conversion, and interpretation of Target.\\
Integration / use of prior assistance & At turn 2, the student reuses the orient parameter introduced in the preceding AI response, providing observable evidence that the AI contribution entered the student's subsequent work. & Subsequent turns introduce further requests, but no student action can be specifically linked to uptake or application of a preceding AI contribution.\\
Evaluation / re-entry after execution & After attempting to execute the preceding solution, the student returns with the resulting error, “AttributeError: module 'pandas' has no attribute 'to\_csv'.” The execution result becomes part of the subsequent interaction, making inspection of the AI-supported work observable. & Across the six turns, the student requests judgments, transformations, and explanations from the AI but does not contribute a comparison, diagnostic claim, criterion, or judgment that makes Evaluation observable.\\
\bottomrule
\end{tabular}
\endgroup{}

}

\end{table}%

A similarly compact sequence illustrates why the coding depends on
relations between turns. The AI introduced JSON export using
\texttt{df.to\_json(...,\ orient=\textquotesingle{}records\textquotesingle{},\ lines=True)}
and identified \texttt{orient} as a parameter controlling the output
structure. The student's next turn, ``to\_json orient types,'' takes up
that AI-introduced element rather than initiating an unrelated request.
The interaction later returns to execution when the student reports
\texttt{AttributeError:\ module\ \textquotesingle{}pandas\textquotesingle{}\ has\ no\ attribute\ \textquotesingle{}to\_csv\textquotesingle{}}.
Read sequentially, these moves provide evidence of uptake and inspection
that would be less informative if the student turns were considered in
isolation.

The contrast is especially informative because the two conversations are
matched on both assessed performance and interaction length. Each
student has five opportunities to respond after an AI contribution, yet
those opportunities unfold differently. In Conversation A, several
re-entry moves provide observable evidence of uptake and inspection of
AI-supported work. In Conversation B, by contrast, the interaction
remains organized mainly around successive directed requests to the
system. The difference in observable process therefore persists even
when score and conversation length are held constant.

The two pairs show that identical assessed performance can coexist with
different observable patterns of participation. In the first pair, both
students display Integration, but Evaluation becomes visible in only one
interaction. In the second, students with the same assignment score and
the same interaction length show different observable configurations:
one interaction contains Direction, Integration, and Evaluation, while
the other contains Direction alone.

These comparisons reinforce the distinction between assessed performance
and observable process. Assignment scores capture how the final work was
evaluated, whereas interaction traces add evidence about how cognitive
work was distributed during AI-mediated activity. This distinction
provides the basis for the discussion that follows on the implications
for educational assessment.

\section{Discussion and final
remarks}\label{discussion-and-final-remarks}

Generative AI weakens a familiar educational inference between assessed
performance and the cognitive process that produced it. Assignment
scores can still indicate whether a student satisfied the criteria of a
task, but they provide limited information about how the relevant
cognitive work was distributed between learner and system. The StudyChat
analysis provides a partial window into that distribution by recovering
observable traces of Direction, Integration, and Evaluation from the
sequential organization of AI-mediated work.

The substantive implication is that similar performance can emerge from
different observable processes. This does not imply that stronger
manifestations of Direction, Integration, or Evaluation should produce
higher scores. Such an expectation would contradict the problem that
motivates the analysis: generative AI can help produce successful
outputs even when substantial parts of the underlying cognitive work
have been delegated. The relevant finding is instead that performance
and observable process provide different kinds of information. Scores
characterize the quality of the assessed product under the criteria of
the assignment; interaction traces reveal aspects of how students
participated in producing it.

This distinction is particularly visible in the StudyChat environment
because programming and data-analysis tasks create recurrent
opportunities for learner re-entry. Code can be executed, results
inspected, discrepancies identified, and procedures revised. These
moments make Integration and Evaluation observable in ways that may be
less readily available in tasks where AI-generated contributions can be
incorporated directly into a final artifact without producing
intermediate consequences. Observability is therefore partly a property
of the task environment rather than simply a property of the learner or
the interaction.

This point has consequences for how epistemic ownership can be studied
in educational settings. Interaction records can provide evidence about
whether students direct the activity, incorporate AI contributions into
subsequent work, and participate in judging those contributions. They
cannot establish epistemic ownership as a whole. In particular,
Answerability requires evidence that the learner can reconstruct and
defend the reasoning embodied in the resulting work. StudyChat does not
systematically elicit that capacity. For the same reason, the present
analysis cannot identify individual cases of justificatory
dispossession. It can establish the empirical condition that makes such
a possibility educationally relevant: successful performance can coexist
with substantially different observable forms of participation in the
cognitive process.

The analysis also points toward a design problem that deserves further
investigation. If interaction traces are informative only when the
activity creates opportunities for relevant forms of participation to
become observable, educational institutions can potentially design
assignments and assessment environments with observability in mind.
Direction is comparatively easy to elicit because students must
ordinarily initiate requests, specify objectives, or choose how to
proceed. Integration and Evaluation may require more deliberate
opportunities for re-entry. An important next question is therefore how
learning activities can be designed so that students are required to do
something with AI contributions rather than merely receive them: apply
them, compare them with alternatives, inspect their consequences,
diagnose problems, justify revisions, or decide whether they should be
retained. The objective would not be to prescribe a single interaction
sequence, but to create conditions under which cognitively relevant
participation leaves evidence that can be examined.

A fuller assessment of epistemic ownership would extend this logic to
Answerability. Product evidence can indicate what was accomplished;
process evidence can illuminate how human and artificial contributions
were organized during production; and subsequent elicitation can examine
whether the learner remains able to reconstruct, defend, or transfer the
reasoning embodied in the work \citep{ajjawi2024authenticity}. How these
sources should be combined, and which forms of elicitation are
appropriate for different learning objectives, remains an open problem
for AI-mediated assessment.

The present analysis has clear limits. StudyChat records only activity
that occurs within the interaction, so the absence of Direction,
Integration, or Evaluation in the transcript cannot be interpreted as
absence of the corresponding cognitive activity. The computational
setting also produces unusually visible feedback and may therefore make
some forms of participation easier to observe than in other domains.
Finally, assignment scores measure assessed performance within a course
rather than learning or epistemic ownership. The analysis is
consequently descriptive: it establishes variation in observable
cognitive participation alongside performance, rather than estimating a
causal relationship between them.

These limitations define the next empirical agenda rather than weakening
the distinction developed here. Generative AI increases the amount of
cognitive work that can occur outside the learner while leaving the
final product largely unchanged as the principal object of assessment.
The resulting challenge is to determine which forms of human
participation remain educationally consequential and how learning
environments can make them observable. Epistemic ownership provides a
framework for organizing that problem. Its contribution is to direct
attention toward the learner's continuing command over AI-mediated work
and toward the evidence needed to evaluate that command when successful
performance alone can no longer reveal how the work was produced.

\bibliography{PrePrint_Fabrega_Same_Performance_Different_Process_literature.bib}

\end{document}